\documentclass[aps,prl,showpacs,twocolumn]{revtex4}
\usepackage{graphicx}

\newcommand{\ket}[1]{\left| #1\right\rangle}

\newcommand{\ketbra}[2]{\left| #1\right\rangle\!\left\langle#2\right|}

\begin{document}                

\title{Homodyne tomography characterization and nonlocality of a dual-mode optical qubit}

\author{S. A. Babichev, J. Appel, A. I. Lvovsky \cite{Lvovsky}}

\affiliation{Fachbereich Physik, Universit\"at Konstanz, D-78457
Konstanz, Germany}

\date{\today}

\begin{abstract}
A single photon, delocalized over two optical modes, is
characterized by means of quantum homodyne tomography. The
reconstructed four-dimensional density matrix extends over the
entire Hilbert space and thus reveals, for the first time,
complete information about the dual-rail optical quantum bit as a
state of the electromagnetic field. The experimental data violate
the Bell inequality albeit with a loophole similar to the
detection loophole in photon counting experiments.
\end{abstract}

\maketitle


According to Feynman, single-particle interference is ``a
phenomenon which is impossible, \emph{absolutely} impossible, to
explain in any classical way, and which has it in the heart of
quantum mechanics" \cite{FLF}. The explanation offered by quantum
mechanics is that a particle incident onto a beam splitter is not,
as classically expected, randomly reflected or transmitted, but
forms a \emph {delocalized} coherent superposition of these two
possibilities. The quantum state in the two beam splitter output
modes  $A$ and $B$ can be expressed as
\begin{equation}\label{psimin}
\ket{\Psi}=\tau\ket{1_A,0_B}-\rho\ket{0_A,1_B}, \end{equation}
where $\tau^2$ and $\rho^2$ are, respectively, the beam splitter
transmission and reflectivity, and the kets in the right-hand side
are written in the particle-number basis.



In this article, we investigate the delocalized state formed by a
photon. We employ homodyne tomography, a state characterization
technique based on phase-sensitive measurements of the
electromagnetic field quadratures. Although there is only one
light particle, we find it to affect the field in both modes
simultaneously, giving rise to nonclassically correlated,
phase-dependent quadrature statistics. This is a direct
consequence of the coherent, entangled nature of the state
(\ref{psimin}) and is remarkable because the optical phase in both
the single-photon and vacuum states considered individually is
completely uncertain.

We show that the delocalization of the photon in the state
(\ref{psimin}) can be formulated as a noncontextual violation of
local realism, namely as a violation of Bell's inequality
\cite{Bell}. Nonlocality of the single photon has been discussed
earlier in a number of theoretical publications, and various
experiments were proposed for its demonstration \cite{NLPhoton}
none of which have so far been realized. The present work achieves
this goal by using a very different measurement method and by
introducing an assumption which resembles the fair sampling
assumption in traditional experiments on nonlocality \cite{ExpBI}.
To our knowledge, this is the first experimental verification of
Bell's theorem performed in a continuous-variable setting.

Apart from its fundamental implications, our experiment finds its
use in the linear optical implementation of quantum optical
information processing \cite{klm}, in which the state
(\ref{psimin}) plays the role of a quantum bit. To date,
characterization of optical qubits has been based on studying
relative photon number statistics in each mode and in their
various linear superpositions as well as (in the case of multiple
qubits) photon number correlations between modes.  Employing this
approach, White {\it et al.} have implemented tomography of
entangled two-qubit systems \cite{White1}. Most recently, this
technique has been extended to characterization of quantum
dynamical processes \cite{ProcessTomo}. Similar methods are used
in celebrated quantum cryptography and quantum teleportation
protocols.

A major drawback of the photon counting approach to quantum state
characterization is the {\it a priori} assumption that the modes
involved are either in one of the states $\ket{1_A,0_B}$,
$\ket{0_A,1_B}$ or in their linear combination. As a result, being
characterized are not the quantum states of the carrier modes, but
their \emph{projections} onto a subspace spanned by these basis
vectors. It is neglected that in actual experimental situations,
most of the time the state of the modes does not belong to this
subspace but is the double vacuum $\ket{0_A,0_B}$. These
overwhelming events, as well as all other events not falling into
the qubit frame, were simply eliminated from the analysis.

The homodyne tomography method employed in this paper is free from
this disadvantage as it permits \emph{complete} characterization
of a quantum optical ensemble. So far it has been employed to
study single- \cite{HomoTomoSingle} and multi-mode \cite{Vasilyev}
squeezed states and, more recently, the single-photon and related
states \cite{FockRefs}. Here we apply this technique to the
delocalized photon state (\ref{psimin}) and evaluate its density
matrix and its Wigner function. A detailed theoretical analysis of
different aspects of an experiment such as ours was made by Jacobs
and Knight \cite{JacobsKnight} as well as Grice and Walmsley
\cite{GriceWalmsley}.

In our experiment, the initial single-photon state was prepared by
means of a conditional measurement on a biphoton produced via
parametric down-conversion \cite{Mandel86Gran86}. We used
frequency-doubled 2-ps pulses from a mode-locked Ti:Sapphire laser
running at $\lambda=790$ nm which underwent down-conversion in a
BBO crystal, in a type-I frequency-degenerate configuration
\cite{FockRefs}.

Field quadratures of the delocalized photon ensemble were measured
by means of two homodyne detectors (associated with fictitious
observers Alice and Bob) placed into each beam splitter output
channel (Fig.~1(a)). The local oscillator fields were provided by
the master Ti:Sapphire laser and contained several million photons
per pulse. Their optical mode had to be matched to those of the
state under investigation \cite{MMpaper}. A specially designed
balanced detector employed two Hamamatsu S3883 photodiodes of a
94-\% quantum efficiency whose photocurrents were directly
subtracted \cite{BHD}. The amplification circuit provided
suppression of the electronic noise in a large frequency range
from DC to 2--3 MHz, which enabled time-resolved quadrature
measurements for each pulse. The time-integrated homodyne detector
output directly corresponded to a quadrature value, with a scaling
reference obtained from the vacuum state shot noise.

With every incoming photon, both detectors made a measurement of
the field quadrature $X_A$ and $X_B$ with the local oscillators'
phases set to $\theta_A$ and $\theta_B$, respectively. The
quadrature statistics collected at various phases were used to
reconstruct the density matrix of the generated ensemble.

Generally, homodyne tomography reconstruction of a two-mode state
requires fine control of both local oscillator phases involved.
However, the particular state studied in the present experiment
was generated by splitting a single photon which has no optical
phase. The sum of Alice's and Bob's phases $\theta_A+\theta_B$ is
therefore meaningless and does not affect the homodyne statistics.
Therefore, we let this phase vary randomly and controlled only the
relative phase $\delta \theta=\theta_A-\theta_B$ of the two modes.
During an experimental run, we varied this parameter slowly over a
$2\pi$ range, and acquired a few hundred thousand pairs
$(X_{A},X_{B})$ from the homodyne detectors. We have executed two
data acquisition runs using two different beam splitters with
transmissions $\tau^2$ equal to 0.5 and 0.08.

\begin{figure}[!t]
    \begin{center}
        \includegraphics[width=0.45\textwidth]{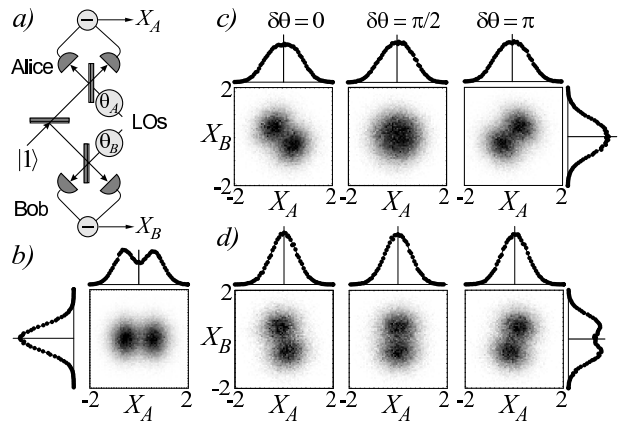}
        \caption{(a) Scheme of the experimental setup. LOs: local oscillators. (b-d)
        Histograms of the experimental quadrature statistics ${\rm
        pr}_{\delta\theta}(X_A,X_B)$ for the ``zero reflectivity" (b), symmetric (c), and
        highly reflective (92\%, d) beam splitters.
        Phase-dependent quadrature correlations are a consequence
        of the entangled nature of the state $\ket{\Psi}$. Also
        shown are individual histograms of the data measured by
        Alice and Bob, which are phase-independent. }
        \label{corr}
    \end{center}
\end{figure}

Fig.~1(c,d) shows histograms of the dual-mode quadrature
measurements. The two-dimensional distribution ${\rm
pr}_{\delta\theta}(X_A,X_B)$ indicates the probability of
detecting a particular pair $(X_A,X_B)$ of quadratures at a given
local oscillator phase setting. These densities are the marginal
distributions of the four-dimensional Wigner function of the
two-mode ensemble being measured. They can be used for its direct
reconstruction via the inverse Radon transformation.

In this work, however, he have used a more precise
maximum-likelihood reconstruction technique \cite{Rehacek01,B5}.
We applied the iterative algorithm described in \cite{MaxLikLvov}
to the two-mode density matrix of the reconstructed ensemble in
the Fock representation:
\begin{equation} \hat{\rho}=\sum_{k,l,m,n=0}^{\infty}\rho_{klmn}\ketbra{k_A,l_B}{m_A,n_B}, \label{rho}\end{equation}
 with the photon
number restricted to five per mode. The positive operator-valued
measure (POVM) operator, used to describe the homodyne detector,
was modified with respect to that of an ideal homodyne detector in
two ways. First, we corrected for the non-perfect detection
efficiency associated with linear optical losses and non-unity
quantum efficiency of the homodyne detector photodiodes. These
losses amounted to $1-\eta_{\rm det}=0.14$ and are described
mathematically in the form of a Bernoulli transformation affecting
every photon entering the detector \cite{leon,B5,MaxLikLvov}.



Secondly, due to a random variation of the phase
$\theta_A+\theta_B$, all the POVM elements, for which $m+n\ne
k+l$, vanish, and so do the respective elements of the
reconstructed density matrix --- akin to the off-diagonal terms of
any phase-randomized single-mode ensemble.

\begin{figure}[!t]
    \begin{center}
        \includegraphics[width=0.4\textwidth]{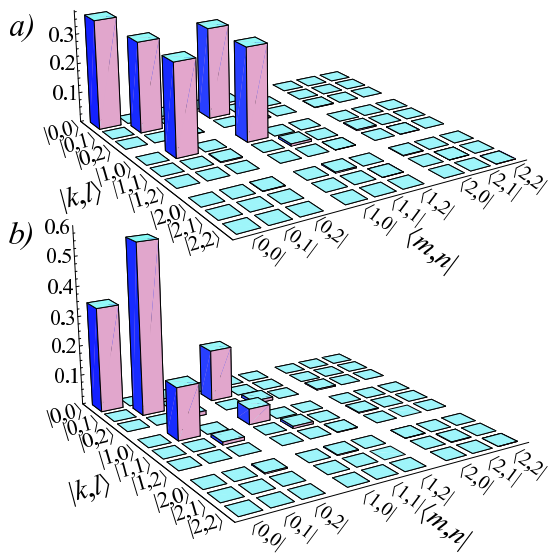}
        \caption{Density matrix (absolute values) of the measured ensemble for
        the symmetric (a) and highly reflective (b) beam splitters in
        the photon number representation (\ref{rho}).
        }
        \label{dmsym}
    \end{center}
\end{figure}

The reconstructed density matrix (Fig.~2) features a strong
contribution of the double-vacuum term $\ketbra{0,0}{0,0}$. This
is a consequence of imperfect preparation of the initial single
photon: instead of the state $\ket{1}$, a statistical mixture
$\hat\rho_{\ket{1}}=\eta\ketbra{1}{1}+(1-\eta)\ketbra{0}{0}$ is
available at the beam splitter input \cite{FockRefs,MMpaper}. The
vacuum fraction is directly transferred to the dual-mode ensemble:
$\hat\rho=\eta\ketbra{\Psi}{\Psi}+(1-\eta)\ketbra{0,0}{0,0}$. The
reconstructed ensemble is in excellent agreement with the above
equation and corresponds to the preparation efficiency
$\eta=0.64$.


Some insight into the quadrature dynamics of the nonlocal
single-photon state can be
gained by analyzing its Wigner function (WF). 
Theoretically, the transformation of a
quantum ensemble by a beam splitter can be expressed as a rotation
of its WF in the four-dimensional phase space \cite{leon}:
\begin{eqnarray} &&W(X_A,P_A,X_B,P_B)=W_{in}(\tau X_A+\rho X_B,\\
\nonumber&&\quad\quad\quad \tau P_A+\rho P_B,\tau X_B-\rho
X_A,\tau P_B-\rho P_A).\end{eqnarray} In our experiment, the input
WF is an uncorrelated product of the Wigner functions of the
single-photon and vacuum states in the beam splitter input modes:
\begin{equation}
W_{in}(X_1,P_1,X_2,P_2)=W_{\ket{1}}(X_1,P_1)\times
W_{\ket{0}}(X_2,P_2).
\end{equation} A rotation in the phase space entangles the input
mode, generating the ensemble $\ket{\Psi}$. The effect of this
rotation can be seen in Fig.~3 which shows the cross-sections of
the four-dimensional WF of the reconstructed ensemble through two
different planes. As expected for the single photon, the WF is
negative in the phase space origin, albeit the negativity is
reduced due to a non-perfect preparation efficiency.

\begin{figure}[!t]
    \begin{center}
        \includegraphics[width=0.4\textwidth]{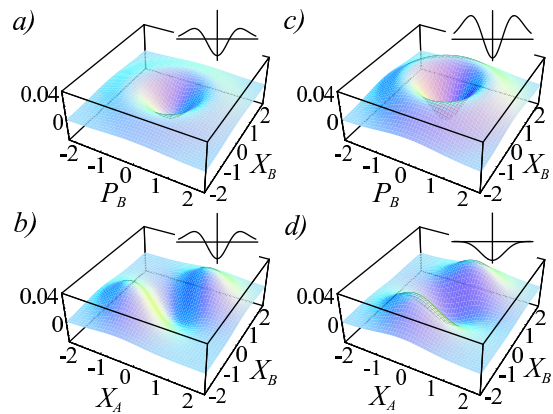}
        \caption{Cross-sections of the two-mode Wigner function
        $W(X_A,X_B,P_A,P_B)$ of the reconstructed ensemble for the
        symmetric (left column) and highly reflective (right
        column) beam splitters. (a, b): cutting plane
        $(X_A=P_A=0)$; the WF is axially symmetric. (c, d):
        cross-sections through the plane $(P_A=P_B=0)$ possesses
        mirror symmetry with symmetry planes oriented at an angle
        $\tan^{-1}(\rho/\tau)$ with respect to the coordinate
        axes. Insets: cross-sections through $X_B=0$.
        }
        \label{wig}
    \end{center}
\end{figure}
A similar argument can be used to understand the shape of the
measured quadrature probability densities. Consider an experiment
in which the beam splitter has zero reflectivity, so that Alice
measures the field quadratures of the single-photon state, and
Bob, at the same time, those of the vacuum state. The joint
quadrature statistics are, naturally, uncorrelated [Fig.~1(b)].
However, if the two modes are subjected to a beam splitter
transformation prior to their measurement, their field amplitudes
will mix, the quadrature histogram will rotate by an angle
$\tan^{-1}(\tau/\rho)$ and become correlated [Fig.~1(c,d)]:
\begin{eqnarray}\label{bseffect}
  X_{A}&=&\tau X_1+\rho X_2\\
  \nonumber X_{B}&=&-\rho X_1+ \tau X_2.
\end{eqnarray}

The above relation is valid only if Alice and Bob measure the same
quadrature of the incoming fields, i.e. for $\delta\theta=0$ or
$\delta\theta=\pi$. On the other hand, the setting
$\delta\theta=\pi/2$ corresponds to a measurement of orthogonal
quadratures, e.g. $X$ by Alice and $P$ by Bob. It can be
interpreted as a simultaneous, unprecise measurement of the
canonically conjugated position and momentum quadratures of the
input single-photon state with an eight-port homodyne detector
\cite{WalkerCaroll,leon}. Furthermore, the distribution ${\rm
pr}_{\pi/2}(X_A,X_B)$ acquired with a symmetric beam splitter
[Fig.~1(c), middle] is equal to the $Q$-function of this state
\cite{LeonPaul}.


As discussed above, these phase-dependent features of the
quadrature statistics are incompatible with a naive picture
according to which the photon is randomly localized in one of the
beam splitter output channels. The nonlocal character of our
measurements can be formalized as a violation of the Bell
inequality \cite{Bell}. In order to apply the Bell theorem to the
continuous field quadratures measured in this experiment, the
latter had to be converted to a dichotomic format. We proceeded as
follows. Suppose the output of each homodyne detector is processed
by a discriminator producing the value $S=1$ if the observed
quadrature exceeds a certain threshold $T$, $S=-1$ if it is below
$-T$, and generating no output otherwise. Restricting to the
events in which both discriminators have generated a value, we
studied correlations $E_{AB}=<S_A S_B>$ between the values
acquired by Alice and Bob.

\begin{figure}
    \begin{center}
        \includegraphics[width=0.45\textwidth]{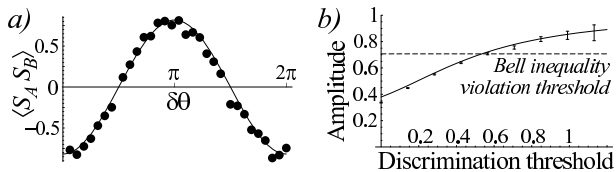}
        \caption{Bell inequality violation by a threshold-selected subset of the experimental
        data [symmetric beam splitter]. (a) Interference pattern for the threshold of
        $T=0.85$ shows an amplitude of $0.818\pm0.018$; (b)
        theoretical and experimental
        dependence of the amplitude on the threshold value.
        A higher threshold increases the amplitude, but also reduces
        the number of selected data.
        }
        \label{quad}
    \end{center}
\end{figure}
The correlation $E_{AB}$ as a function of $\theta_A-\theta_B$
exhibits quantum interference [Fig.~4] which is similar to that
observed in Bell's Gedankenexperiment for the spin correlation as
a function of the angle between the axes of the Stern-Gerlach
apparata. Increasing the threshold $T$ leads to a higher
visibility of the interference pattern. For $T>0.54$, the
amplitude exceeds $1/\sqrt{2}$ and the Bell inequality is
violated. Interpreting our data in this manner, we observed
violations by up to six standard deviations.

One might object that such an interpretation involves rejection of
some of the the acquired data. This objection is valid; however,
similar selection is also a part of the traditional way of
verifying Bell's theorem by studying the counting statistics of a
polarization-entangled photon pair \cite{ExpBI}. Indeed, every
single-photon counter has a discriminator as a part of its
electronic circuit, and the signals not exceeding the
discrimination threshold are disregarded. Viewed outside of a
context of any particular physical theory, both tests possess an
equal degree of validity: they both involve a fair sampling
assumption and thus suffer from the detection loophole
\cite{detloop}.

Is our \emph{entire} data set noncontextually nonlocal? Although a
number of schemes of reformulating Bell's theorem for continuous
variables \cite{bicv} have been proposed, we have not yet found
one that could be applied to our original data. On the other hand,
we could not come up with a local hidden variable model that would
replicate the quadrature statistics observed in this experiment
\cite{Wigner55}.

In summary, we have characterized a two-mode optical qubit using
homodyne tomography, reconstructing its density matrix and the
Wigner function. For the first time, complete information about
this quantum ensemble is revealed, including those terms that do
not belong to the qubit subspace of the Hilbert space. The
experimental data demonstrate a nonlocal character of the
delocalized single photon but the question of noncontextual nature
of this nonlocality remains open.

We are grateful to A. M. Steinberg, who asked about a possibility
of selecting a data subset that violates Bell's inequality, and to
J. Ries and B. Brezger for helpful discussions. We are supported
by the Deutsche Forschungsgemeinschaft and the Optik-Zentrum
Konstanz.

\end{document}